\documentstyle[prl,aps,floats,epsf]{revtex}

\begin{document}
\draft

\twocolumn[\hsize\textwidth\columnwidth\hsize\csname
@twocolumnfalse\endcsname

\title {Doping Dependent Density of States and Pseudogap 
Behavior in La$_{2-x}$Sr$_x$CuO$_4$}
\author{A. Ino, T. Mizokawa , K. Kobayashi, A. Fujimori}
\address{Department of Physics, University of Tokyo, Bunkyo-ku, Tokyo 
113, Japan}
\author{T. Sasagawa, T. Kimura\cite{adr2}, K. Kishio, K. 
Tamasaku\cite{adr1}, H. Eisaki, S. Uchida}
\address{Department of Superconductivity, University of Tokyo, 
Bunkyo-ku, Tokyo 113, Japan}
\maketitle

\begin{abstract}
We have made a high-resolution photoemission study of 
La$_{2-x}$Sr$_x$CuO$_4$ in a wide hole concentration ($x$) range from 
a heavily overdoped metal to an undoped insulator.  As $x$ decreases, 
the spectral density of states at the chemical potential ($\mu$) is 
suppressed with an $x$-dependence similar to the suppression of the 
electronic specific heat coefficient.  In the underdoped region, the 
spectra show a pseudogap structure on the energy scale of 0.1 eV. The 
width of the pseudogap increases with decreasing $x$ following the 
$x$-dependence of the characteristic temperatures of the magnetic 
susceptibility and the Hall coefficient.
\end{abstract}

\pacs{PACS numbers: 71.30.+h, 74.72.Dn, 74.25.Jb, 79.60.Bm}

\vskip1pc]

\narrowtext

In order to understand the mechanism of high-temperature 
superconductivity in doped cuprates, a central issue has been the 
evolution of the electronic structure with hole doping near the 
filling-control metal-insulator transition (MIT).  In spite of 
extensive photoemission studies, it still remains unclear how the 
electronic structure evolves, especially, between underdoped metal and 
antiferromagnetic insulator.  For a systematic study of the doping 
dependence near the MIT, La$_{2-x}$Sr$_x$CuO$_4$ (LSCO) is a suitable 
system.  It has the simplest crystal structure with single CuO$_2$ 
layers and the hole concentration in the CuO$_2$ plane is well 
controlled over a wide range and uniquely determined by the Sr 
concentration $x$ (and small oxygen non-stoichiometry).  So far 
photoemission studies of high-$T_{c}$ cuprates were concentrated on 
Bi$_2$Sr$_2$CaCu$_2$O$_8$ (BSCCO) and YBa$_2$Cu$_3$O$_{7-y}$ (YBCO) 
systems.  With the LSCO system, one can investigate the electronic 
structure of the CuO$_2$ plane continuously from the heavily overdoped 
limit ($x\sim0.35$) to the undoped insulator ($x=0$) in a single 
system.

Recently, in underdoped cuprates a ``normal-state gap'' behavior above 
$T_c$ has been observed by angle-resolved photoemission spectroscopy 
(ARPES) in BSCCO \cite{ARPES} and a ``spin-gap behavior'' by NMR in 
YBCO \cite{Yasuoka}.  The magnitude of the normal-state gap is of the 
same order as the superconducting gap at optimal doping.  Meanwhile, 
underdoped cuprates have characteristic temperatures which are 
considerably {\it higher} than $T_{c}$ in the uniform magnetic 
susceptibility \cite{chi}, the electronic specific heat \cite{Loram}, 
the Hall coefficient \cite{Hwang} and the electrical resistivity 
\cite{chi}.  All these characteristic temperatures show similar 
behaviors in LSCO: they increase from $\sim$300 K at optimal doping 
$x\sim0.15$ to $\sim$600 K at $x\sim0.1$ for the LSCO system, 
suggesting a pseudogap-type electronic structure.  In addition, it can 
be reconciled only if a pseudogap is opened at the chemical potential 
that both the electronic specific heat coefficient $\gamma$ 
\cite{Loram,Momono} and the chemical potential shift with doping 
\cite{chempot} are suppressed towards the MIT. To obtain a full 
picture of the evolution of those ``gaps,'' it is also necessary to 
know the total density of state (DOS), which is most directly observed 
by angle-integrated photoemission spectroscopy (AIPES).  In the 
present study, we have performed high-resolution AIPES measurements 
and made a systematic study of the doping dependence of the electronic 
structure of the LSCO system, focusing on the evolution of a 
pseudogap, in a wide hole concentration range from $x=0$ (undoped) to 
$x=0.3$ (heavily overdoped).

High-quality single crystals of LSCO were grown by the 
traveling-solvent floating-zone method.  The samples had $T_c$'s of 
18, 34, 32, 21 and 0 K for $x = 0.074, 0.13, 0.175, 0.203$ and 0.30, 
respectively, with transition widths of about 0.5 K. All the samples 
were annealed to make the oxygen content stoichiometric.  The $x=0$ 
sample was annealed in a reducing atmosphere (100 Torr O$_2$ at 
800$^\circ$C) and confirmed that the N\'eel temperature was higher 
than 250 K, meaning that the hole concentration was less than 0.004 
\cite{CYchen}.  Photoemission measurements were carried out using the 
He {\footnotesize I} line ($h\nu=21.4$ eV) with an overall energy 
resolution of $\sim22$ meV. The base pressure in the spectrometer was 
in the $10^{-11}$ Torr range.  Clean surfaces were obtained by {\it in 
situ} scraping with a diamond file in every 40 minutes.  In order to 
minimize the degradation of sample surfaces, the measurements were 
performed at $T\sim18$ K. Only $x=0$ sample was measured at $\sim70$ K 
to avoid charging effect, but still a slight charging effect (at most 
$\sim10$ meV) could not be eliminated.  Energies were carefully 
calibrated using Au evaporated on each sample so that its uncertainty 
was about 1 meV. The spectra were corrected for the He {\footnotesize 
I}$^\ast$ satellite assuming that He {\footnotesize I}$^\ast$/He 
{\footnotesize I} ratio is constant in all the spectra.

\begin{figure}
    \center \epsfxsize=68mm \epsfbox{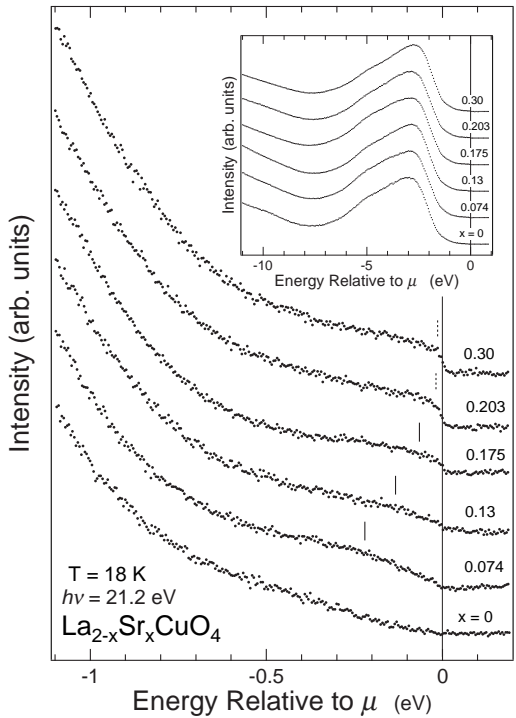} \vspace{0.5pc} 
    \caption{Photoemission spectra of La$_{2-x}$Sr$_x$CuO$_4$ near the 
    chemical potential $\mu$.  The vertical bars mark the point of 
    maximum curvature and represent the energy of the pseudogap.  
    Inset shows the entire valence-band spectra.}
\label{spectra}
\end{figure}

Figure \ref{spectra} shows photoemission spectra for various 
compositions.  The whole valence-band spectra given in the inset show 
no trace of a hump at $\sim-9$ eV, indicating the high sample quality 
and the cleanliness of the sample surfaces.  The spectra have been 
normalized to the maximum intensity of the valence band since the area 
of the valence band was hard to determine accurately due to the 
experimentally ambiguous background.  As $x$ decreases, the intensity 
at the chemical potential $\mu$ decreases and disappears in the 
insulating phase ($x=0$).  While in the overdoped region ($x > 0.2$) 
the spectra show an ordinary metallic Fermi edge, in the underdoped 
region ($x < 0.15$) the spectra show a pseudogap-type lineshape around 
$\mu$ in the sense that the spectral intensity gradually diminishes 
towards $\mu$ from somewhat below $\mu$ as marked by vertical bars in 
Fig.~\ref{spectra}.  Here the vertical bars indicate the points of 
maximum curvature.  It appears that a rather large pseudogap develops 
with decreasing $x$, with its width and depth increasing towards 
$x=0$.  Note that the energy scale of the DOS suppression is as large 
as the order of $\sim0.1$ eV (``high-energy pseudogap'') and therefore 
that it does not correspond to the ``normal-state gap'' which has a 
magnitude similar to the superconducting gap ($\sim25$ meV for BSCCO) 
and has been observed in the ARPES spectra of underdoped BSCCO 
(``low-energy pseudogap'') \cite{ARPES}.  In view of the energy scale, 
the marked feature in the AIPES spectra may rather correspond to the 
broad feature at ($\pi$,0) in the ARPES spectra of underdoped BSCCO 
\cite{Shen&schrieffer}.  Unfortunately, since the present spectra were 
taken at a low temperature ($T\sim18$ K), one cannot judge whether the 
low-energy normal-state gap is opened or not in the underdoped LSCO.

\begin{figure}[!b]
    \center \epsfxsize=60mm \epsfbox{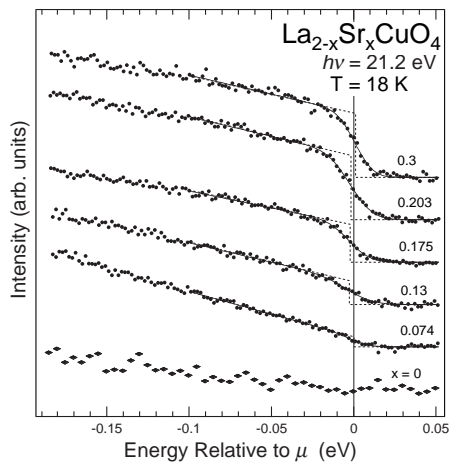} \vspace{0.5pc} 
    \caption{Enlarged view of the spectra near the chemical potential 
    $\mu$.  The spectra have been fitted to a linear DOS multiplied by 
    the Fermi-Dirac distribution function convoluted with a Gaussian 
    representing the instrumental resolution.  Hypothetical $T\to0$ 
    spectra without Gaussian broadening are also shown by dotted 
    lines.}
\label{fitting}
\end{figure}

\begin{figure}[!t]
    \center \epsfxsize=57mm \epsfbox{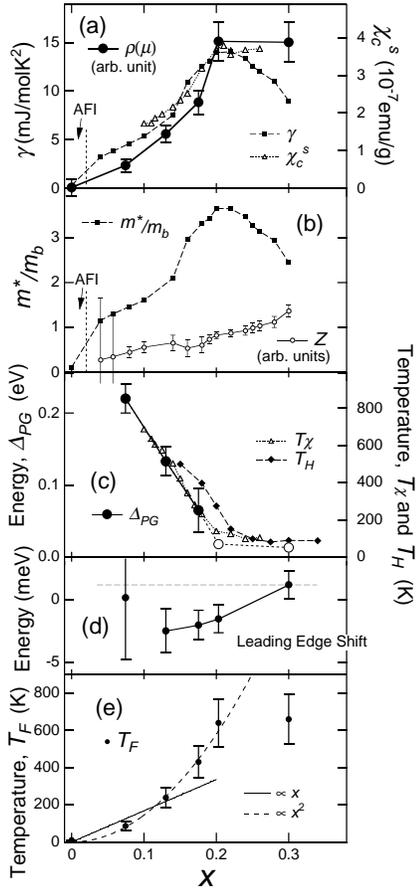} \vspace{3pt} 
    \caption{Doping dependence of (a) the DOS at $\mu$, $\rho(\mu)$, 
    compared with the electronic specific heat coefficient $\gamma$ 
    \protect\cite{Momono} and the Pauli paramagnetic component 
    $\chi_c^s$ of the spin susceptibility \protect\cite{chi}, (b) the 
    renormalization factor $Z=\rho (\mu) / N^\ast(\mu)$ and the mass 
    enhancement factor $m^\ast/m_b=N^\ast(\mu)/ N_b(\mu)$, (c) the 
    pseudogap energy $\Delta_{PG}$ compared with the characteristic 
    temperatures $T_{\chi}$ where the magnetic susceptibility $\chi$ 
    takes a maximum\protect\cite{chi} and $T_H$ below which the Hall 
    coefficient $R_H$ increases \protect\cite{Hwang}, (d) the position 
    of the leading edge and (e) the ``Fermi temperature" (coherence 
    temperature) $T_F = (1/\pi k_B) 
    \rho(\omega)/(\partial\rho(\omega)/\partial\omega)|_{\omega=\mu}$ 
    of doped holes.}
\label{Allin1}
\end{figure}

In order to determine the spectral DOS at $\mu$, $\rho(\mu)$, and the 
precise position of the leading edge, the spectra near $\mu$ ($>-0.1$ 
eV) have been fitted to a linear DOS multiplied by the Fermi-Dirac 
distribution function convoluted with a Gaussian of the instrumental 
resolution, as shown in Fig.~\ref{fitting}.  The obtained $\rho(\mu)$ 
are shown in Fig.~\ref{Allin1}~(a), where error bars include 
uncertainties in the normalization procedure due to subtle changes in 
the valence-band lineshape.  In Fig.~\ref{Allin1}~(a), $\rho(\mu)$ is 
compared with the specific heat coefficient $\gamma$ \cite{Momono} and 
the Pauli-paramagnetic component $\chi_s^c$ of the spin susceptibility 
\cite{chi}.  The three quantities, $\rho(\mu)$, $\gamma$ and 
$\chi_s^c$, show quite similar $x$-dependences: for $x>0.2$, with 
decreasing $x$ they slowly increase or remain nearly constant, take a 
maximum around $x=0.2$ and then decrease towards $x\sim0$ for $x<0.2$.  
A similar behavior was predicted by Hubbard-model calculations 
\cite{Duffy} although the absolute value of the calculated $\rho(\mu)$ 
was much higher than the observed one.  Using the quasiparticle (QP) 
density at $\mu$, $N^\ast(\mu)$, obtained from $\gamma$ [$N^\ast(\mu) = 
\gamma/(\frac{1}{3}\pi^2 k_B^2)$], and the unrenormalized DOS at 
$\mu$, $N_b(\mu)$, calculated by band theory \cite{LSCObandcalc}, one 
can deduce the mass enhancement factor $m^\ast/m_b\equiv 
N^\ast(\mu)/N_b(\mu)$ and the renormalization factor $Z \equiv 
\rho(\mu) / N^\ast(\mu)$ as shown in Fig.~\ref{Allin1}~(b).  
Figures~\ref{Allin1}~(a) and (b) imply that the decrease in $m^\ast$ 
is driven by the decrease in $\rho(\mu)$ associated with the 
development of the high-energy pseudogap.  The renormalization factor 
(or equivalently the QP spectral weight) $Z$ also decreases towards 
$x=0$, but it is not clear whether $Z$ goes to zero or remains finite 
as $x \rightarrow 0$ because of experimental uncertainties.  It should 
be noted that the QP mass is enhanced as $x$ decreases for $x>0.2$, 
i.e., in the region where the pseudogap is absent.  The behavior for 
$x>0.2$ is similar to that of the typical Mott MIT system 
La$_{1-x}$Sr$_x$TiO$_3$, which shows a QP mass ($\propto m^* 
\propto\gamma$) enhancement with decreasing $x$ until the boundary of 
the antiferromagnetic (AF) phase is reached \cite{TiOspecheat}.

The energy of the high-energy pseudogap $\Delta_{PG}$ was defined by 
the binding energy of the point of maximum curvature determined by 
taking the second derivatives (as marked by the vertical bars in 
Fig.~\ref{spectra}).  The $\Delta_{PG}$ values are plotted in 
Fig.~\ref{Allin1}~(c), which shows that $\Delta_{PG}$ follows the 
$x$-dependence of the characteristic temperatures $T_{\chi}$ 
\cite{chi} and $T_H$ \cite{Hwang}.  Here $T_{\chi}$ and $T_H$ are 
temperatures at which the magnetic susceptibility $\chi$ takes a 
maximum and below which the Hall coefficient $R_H$ increases, 
respectively.  It has also been reported that the temperature 
$T_{\rho}$ below which the electrical resistivity $\rho$ deviates 
downward from the linear-$T$ behavior follows $T_{\chi}$ and $T_{H}$ 
\cite{chi}.  The similar $x$-dependences of $\Delta_{PG}$, $T_{\chi}$, 
$T_H$ and $T_{\rho}$ imply that these temperatures are closely related 
to the development of the high-energy pseudogap.  We may therefore 
refer to those characteristic temperatures as the ``pseudogap 
temperature'' $T_{PG}$.  We then find $\Delta_{PG}/k_BT_{PG}\simeq3$, 
indicating the interaction is in the strong coupling regime.

In the low-energy region, the position of the leading-edge midpoint 
obtained from the fitting is shown in Fig.~\ref{Allin1}~(d).  In going 
from the overdoped region to the optimum doping, the edge is shifted 
downward by $\sim 4$ meV. If we attribute the shift to the opening of 
a superconducting gap, we obtain the ratio $2\Delta/k_BT_c\sim2$, 
which is smaller than the typical value ($4-6$) deduced from ARPES of 
BSCCO and YBCO \cite{Dessau}.  However, if the superconducting gap is 
anisotropic as in $d$-wave pairing, it is understandable that the 
angle-integrated spectra fitted to the simple step function give a 
smaller leading-edge shift than the ARPES data.

In the underdoped regime, since $\rho(\mu)$ is small and the slope of 
DOS at $\mu$, $\partial\rho(\omega)/\partial\omega|_{\omega=\mu}$, is 
steep, the Fermi edge is obscured at high temperatures and the 
Fermi-Dirac distribution lose its meaning.  The crossover temperature 
for such a disappearance of the Fermi edge is thus given by $T_F = 
(1/\pi k_B) \rho(\omega)/\frac{\partial\rho(\omega)} 
{\partial\omega}|_{\omega=\mu}$ [Fig.~\ref{Allin1}~(e)] and may be 
called the ``coherence temperature'' of the doped holes.  $T_F$ may 
also be understood as the ``Fermi temperature'' of the doped holes 
because if one linearly extrapolates the DOS $\rho(\omega)$ beyond 
$\mu$, then a hole pocket has the Fermi energy of $\varepsilon_F 
\equiv \pi k_BT_F$.  For such a hole pocket, transport at high 
temperatures $T\gg T_F$ would be dominated by the incoherent charge 
dynamics of the doped holes \cite{Jaklik}.  Figure~\ref{Allin1}~(e) 
shows that $T_F$ becomes very low ($<100$K) in the underdoped regime, 
indicating that the thermodynamic and transport properties behave as 
those in ``incoherent metals.'' Note that $T_F$ is considerably 
lower than $T_{PG}$ and even lower than $J/\pi k_B$ in the underdoped 
region.  Under such a condition, the kinetic energy gain of the doped 
holes alone may be insufficient to destroy the AF order, and 
alternatively the disappearance of the long-range order may be 
attributed to strong quantum fluctuations characteristic of 
two-dimensional systems.  Figure~\ref{Allin1}~(e) also suggests that 
for small $x$, $T_F$ scales with $x^2$ rather than $x$.  According to 
the hyperscaling hypothesis of MIT \cite{Imada}, the critical behavior 
of $T_F$ and chemical potential shift $\Delta\mu$ near MIT is given by 
$T_F\propto x^{z/d}$ and $\Delta\mu \propto x^{z/d}$, where $z$ is the 
dynamical exponent of the MIT and $d$ is the spatial dimension 
($d=2$).  Thus the present result ($T_F\propto x^2$) implies $z=4$ and 
is consistent with the observed suppression of the chemical potential 
shift ($\Delta\mu \propto x^{2}$) in underdoped LSCO \cite{chempot}.  
This $z$ value is distinctly different from that of an ordinary 
metal-to-band insulator transitions, where $z=2$.

Finally, let us discuss the microscopic origin of the pseudogap 
behavior.  Since the energy scale of the present pseudogap is of the 
order of the super-exchange energy $J\sim0.1$ eV, it is tempting to 
associate the pseudogap with the development of AF correlations or 
short-range AF order in the underdoped region \cite{Schrieffer,Pines}.  
In this scenario, when the temperature is lowered below $T_{PG}$, the 
AF correlation length increases and consequently the DOS shows a 
pseudogap reminiscent of the AF band gap \cite{Preuss}.  The observed 
pseudogap may also be related with the short-range stripe order 
because it also originates from AF correlations.  It has been stressed 
that LSCO is close to the instability of stripe order even in the 
metallic phase \cite{Emery,Tranquada}.  Since $T_F \ll T_{PG}\sim 
J/\pi k_B$ in the underdoped region, the system cannot be regarded as 
a normal Fermi liquid which is weakly perturbed by AF correlation but 
rather as an AF state perturbed by the motion of doped holes.  As an 
alternative scenario, a pseudogap may be produced by preformed Cooper 
pairs, which lose their coherence above $T_{c}$ but still keep local 
pairing, \cite{Trivedi,Kivelson,Imada2}, or by spinon pairing which 
already occurs above $T_c$ \cite{Fukuyama}.  However, interactions 
which lead to such pairing would have only small energies; that is, 
the energy scale of such a pseudogap should be of the same order as 
that of the superconducting gap ($\Delta_{SC}= 10 -20$ meV) at least 
in the slightly underdoped region.  Probably, such a low-energy 
pseudogap corresponds to the spin gap in NMR of YBCO, the normal-state 
gap in ARPES of BSCCO and the gap observed by tunneling spectroscopy 
and is related with the drop in the electric resistivity just above 
$T_c$ \cite{Tunnel}.  On the other hand, the high-energy pseudogap 
observed in the present work corresponds to the anomalies in the 
magnetic susceptibility, the Hall effect and the specific heat.  Then 
the question may arise whether the high-energy pseudogap and the 
low-energy normal-state gap or spin gap are interrelated or not.  
Further studies are necessary to resolve this interesting issue, which 
is directly related to the mechanism of superconductivity in the doped 
cuprates.

In conclusion, we have found a large pseudogap on the energy scale of 
0.1 eV in the DOS of La$_{2-x}$Sr$_x$CuO$_4$ by AIPES. The presented 
spectra have given us a clear view of the evolution of the pseudogap 
from the overdoped metal to the undoped insulator.  Its evolution with 
decreasing $x$ is represented by the increase of the pseudogap energy 
$\Delta_{PG}$ and the suppression of the spectral DOS $\rho(\mu)$ as 
well as the QP density at $\mu$ near the AF insulating phase.  
Consequently, the coherence temperature $T_F$ is much smaller than 
$\Delta_{PG}$ in the underdoped region, suggesting that an incoherent 
metallic state may dominate the thermodynamic and transport properties 
of the underdoped cuprates.  The origin of the present large pseudogap 
is most likely due to AF correlations or short-range AF order, while 
its relation to the smaller ``normal-state gap'' or ``spin gap'' 
remains an open question and should be addressed in future studies.

We would like to thank M.~Imada and Z.-X.~Shen for helpful 
discussions.  This work is supported by a Grant-in-Aid for Scientific 
Research from the Ministry of Education, Science, Sports and Culture, 
the New Energy and Industrial Technology Development Organization 
(NEDO) and Special Promotion Funds of the Science and Technology 
Agency, Japan.

\end{document}